# Hybrid Magneto Photonic Material Structure for Plasmon Assisted Magnetic Switching

**Alan Hwader Chu[1], Bradlee Beauchamp[1,3], Deesha Shah[2,3], Aveek Dutta[2,3], Alexandra Boltasseva[1,2,3], Vladimir M. Shalaev[2,3], and Ernesto E. Marinero[1,2,3,*]**

[1]*School of Materials Engineering, Purdue University, 701 West Stadium Avenue, West Lafayette, IN 47907, USA*
[2]*School of Electrical and Computer Engineering, Purdue University, 465 Northwestern Ave, West Lafayette, IN 47907, USA*
[3]*Birck Nanotechnology Center, Purdue University, 1205 W State St, West Lafayette, IN 47907, USA*
*[\*]eemarinero@purdue.edu*

**Abstract:**

We have proposed the use of surface plasmon resonances at the interface of hybrid magneto-photonic heterostructures [Opt. Mat. Exp., **7**, 4316 (2017)] for all-optical control of the macroscopic spin orientation in nanostructures in fs time scales. This requires strong spin-photon coupling for the resonant enhancement of opto-magnetic fields, generated through the inverse Faraday effect, in magnetic nanostructures with perpendicular anisotropy. Here we report on the development of nm thick interlayers to control the growth orientation of hcp-Co alloys grown on refractory plasmonic materials to align the magnetic axis out-of-plane, thereby meeting key requirements for the realization of ultrafast magneto-photonic devices.

## 1. Introduction

Since the experimental demonstration of all-optical sub-picosecond demagnetization in metallic nickel films in 1996[1], ultrafast magnetization switching dynamics has been extensively investigated as it offers the possibility for magnetization reversal in fs time scales. All-optical switching (AOS) of magnetization exceeds by several orders of magnitude the precessional switching times of magnetization reversal resulting from applied magnetic fields and spin currents. Faster switching times are needed for spintronic devices to be attractive for beyond-CMOS electronics. 40-fs circularly polarized (CP) laser pulses have been used to reversibly switch the magnetization of by changing the light helicity in a wide range of magnetic materials[2-6]. The interaction between the electric field of CP light and the magnetization, through the Inverse Faraday Effect, generates large opto-magnetic fields (OMF) whose orientation is controlled by the light chirality. To utilize AOS to selectively switch nanomagnets with dimensions over one order of magnitude smaller than the diffraction limit of light, as needed in high-density magnetic memories, it is necessary to confine light in the nanoscale[7-9]. The generation of surface plasmon polaritons has enabled plasmonic applications in nanophotonics beyond the diffraction limit[9-11]. In addition, in recent years the coupling between plasmonic and magnetic effects has been demonstrated to enhance magneto-optical effects[10]. In a recent publication, we have proposed magneto-plasmonic interactions in hybrid magnetic-photonic structures as a vehicle for utilizing AOS to control the magnetization orientation in nanomagnets[11]. CMOS-processing compatible plasmonic materials and nanomagnets with strong perpendicular magnetic anisotropy (PMA) are needed to realize this concept. Refractory plasmonic materials such as TiN and ZrN as reported by Guler *et al.*[12-14] exhibit comparable photonic properties as the noble metals, are mechanically and thermally robust and are CMOS-fabrication compatible. Magnetic materials with strong PMA are the materials of choice for high density magnetic recording. In particular hcp-CoCrPt alloys exhibit superior magnetocrystalline anisotropy that permits the growth of >10nm thick films with strong PMA. As an example, 6nm thick $Co_{70}Cr_{18}Pt_{12}$ films were employed to fabricate 20nm diameter islands for demonstration of magnetic storage recording at densities in excess of 1Tb($10^{12}$ bits)/in[7, 15, 16]. As described by Marinero *et al.*[17], to achieve such ultra-high recording densities, the polar magnetic axis dispersion needs to be small. To this effect, CoCrPt was grown on a plurality of underlayers comprising NiTa/NiW/Ru/TaO$_x$ thin film stacks (40nm total thickness) and c-axis dispersion of only 2.4 degrees was attained [16]. We note that for the case of the hybrid magneto-photonic structures proposed by Dutta *et al.*[11] for ultrafast magnetic switching, said metallic interlayer thicknesses between the plasmonic and magnetic material would introduce significant optical losses, precluding the strong photon-spin coupling required for generating enhanced OMF. To circumvent this problem, in this work we report on the synthesis of few atomic-thick interlayers capable of seeding PMA in CoCrPt thin films grown on plasmonic TiN. Structural and magnetic property measurements are employed to characterize the hybrid structures.

## 2. Materials growth and characterization

### 2.1 Materials growth

Epitaxial TiN (001) exhibits plasmonic properties comparable to Au and is thermally and mechanically stable at high temperatures [14, 18-20]. It can be grown epitaxially onto c-sapphire[18, 21] Si[22], and MgO (001)[23] substrates. In this work, we employ reactive sputtering of Ti in Ar + N$_2$ atmospheres to grow epitaxial films on substrates held at 800°C[14, 20]. The TiN samples upon cooling to room temperature are extracted from the nitride-growth chamber and transferred into a 4-target magnetron sputtering tool. Prior to introducing the MgO/TiN samples into the magnetic-growth chamber the samples were cleaned with acetone, methanol, and isopropyl alcohol (in that order) by using sonication for 5 minutes in each solvent and dried with nitrogen gas. To identify suitable nanoscale interlayers for seeding PMA in CoCrPt grown on TiN, we first investigated the processing parameters and materials requirements for their growth on conventional Si/SiO$_2$ substrates. Consistent with the work of Platt *et al.*[24], $Co_{70}Cr_{18}Pt_{12}$ thin films(1 - 10nm thick) with strong PMA, were grown on Ta(5nm)/Ru(10nm) dual seed layers. Their magnetic and structural properties were used to benchmark the properties of equivalent thickness $Co_{70}Cr_{18}Pt_{12}$ alloys grown on both Si/SiO$_2$ and TiN substrates coated with seed layers of different material compositions and thicknesses. All depositions of the magnetic and the seed layers were done without substrate heating in pure Ar atmospheres.

### 2.2 TiN/CoCrPt magneto-photonic structure growth

Compared to $Co_{20}Fe_{60}B_{20}$, the most commonly employed material in magnetic tunnel junctions (MTJ)[25], the ternary alloy $Co_{70}Cr_{18}Pt_{12}$ has a lower saturation magnetization, higher magnetocrystalline anisotropy and coercivity[26]. The magnetic properties of CoCrPt films make them attractive for lowering the switching current and improving the thermal stability of MTJ[27]. $Co_{70}Cr_{18}Pt_{12}$ thin films were deposited at ambient temperature onto oxidized silicon substrates: Si(001)/SiO$_2$(300nm) as well as onto MgO(001)/TiN(30nm) substrates by DC magnetron sputtering in a four-target sputter chamber with a base pressure of < $10^{-7}$ Torr (PVD Products, Inc.). TiN layers were grown epitaxially along the [001] direction at 800°C using reactive sputtering in a separate chamber with a base pressure of < $10^{-7}$ Torr (PVD Products, Inc.). All the substrates were cleaned in acetone and methanol and then dried in nitrogen gas before deposition. We note that when TiN is exposed to air, a few nm thick surface layers of TiO$_x$N$_y$ are known to form[20]. They can be expected to influence the nucleation and growth of subsequently deposited materials. All seed and underlayers and combinations thereof (Ta, Ru, Ta/Ru, CoCrPtTa alloys) investigated were grown in-situ in the magnetic deposition chamber prior to depositing the magnetic material. Their thicknesses ranged from 1-15 nm. Ta and Ru films were grown with an Ar sputter pressure of 1.8mTorr, whereas a pressure of 3mTorr was employed for the growth of CoCrPt and the co-sputtered CoCrPtTa seed layers. Materials deposition rates were calibrated via atomic force microscopy (AFM). All thin film stacks were overcoated with either Ru (5nm) or Ta (10nm) to prevent oxidation and concomitant stoichiometry changes of the CoCrPt alloy.

*2.3 Characterization*

Crystallographic properties were investigated using a Panalytical X-ray diffractometer with Cu-K$\alpha_1$ radiation equipped with mirror optics for signal to noise improvements. To investigate the degree of crystallographic texturing, rocking measurements were performed using a slit acceptance angle of 0.27° placed in front of the detector. For CoCrPt, the full-width-half-maximum (FWHM) of the rocking curve measurements of the (0002) reflection is an indicator of the degree of alignment of the c-axis out-of-plane and thus, of the magnetic axis orientation perpendicular to the thin film plane. For these measurements, the incident beam and the detector alignment were fixed at the center of the Bragg peak angle of interest. During the measurement, the sample holder is slightly tilted and a scan rate of 0.01 degree/step and 4 sec/step was employed to improve the signal/noise ratio. The magnetic properties of the samples were measured with a Quantum Design MPMS-3 superconducting quantum interference device (SQUID) magnetometer with $10^{-8}$ emu sensitivity. The measurements were conducted at room temperatures with applied magnetic fields ranging from -2T to +2T. Saturation magnetization (M$_S$) is attained for applied fields of ~1T. The remanence magnetization (M$_R$) is measured in the absence of an applied magnetic field. We use the following figures of merit (FOM) to characterize the magnetic properties of the Co$_{70}$Cr$_{18}$Pt$_{12}$ films grown on (Co$_{70}$Cr$_{18}$Pt$_{12}$)$_x$Ta$_y$ seed layers: the ratio of the out-of-plane to in-plane remanent magnetization (**M$_R^{OP}$/M$_R^{IP}$**) and the ratio of the out-of-plane remanent magnetization over that measured at saturation, the squareness (**S\*=M$_R^{OP}$/M$_S^{OP}$**). These FOM are compared and discussed with reference to those obtained when Co$_{70}$Cr$_{18}$Pt$_{12}$ is grown on Ta(5nm)/Ru(10nm) dual layers. Ellipsometry characterization of trilayer CoCrPtTa/CoCrPt/Ru metallic structures grown on TiN were conducted utilizing variable angle spectroscopic ellipsometry at angles of 50° and 70° for wavelengths from 500 nm to 2000 nm. Details of the measurements and results are provided in the supplementary information.

## 3. Results

*3.1 Co$_{70}$Cr$_{18}$Pt$_{12}$ thin film growth on Si/SiO$_2$ and MgO/TiN substrates*

The magnetic hysteresis loops of Co$_{70}$Cr$_{18}$Pt$_{12}$ thin films grown directly on Si/SiO$_2$ and MgO/TiN(30nm) are shown in Fig. 1. Their magnetization orientation lies parallel to the thin film plane. This is evidenced in Table 1 from the low value of M$_R^{OP}$/M$_R^{IP}$. Strong PMA is evident in a magnetic film when S\* ~ 1.0 and M$_R^{OP}$/M$_R^{IP}$ becomes very large. The M$_R^{OP}$/M$_R^{IP}$ values of Co$_{70}$Cr$_{18}$Pt$_{12}$ films grown directly on Si/SiO$_2$ and MgO/TiN are 0.16, and 0.11, respectively. Indicating that these films have their magnetic axis preferentially aligned in the thin film plane.

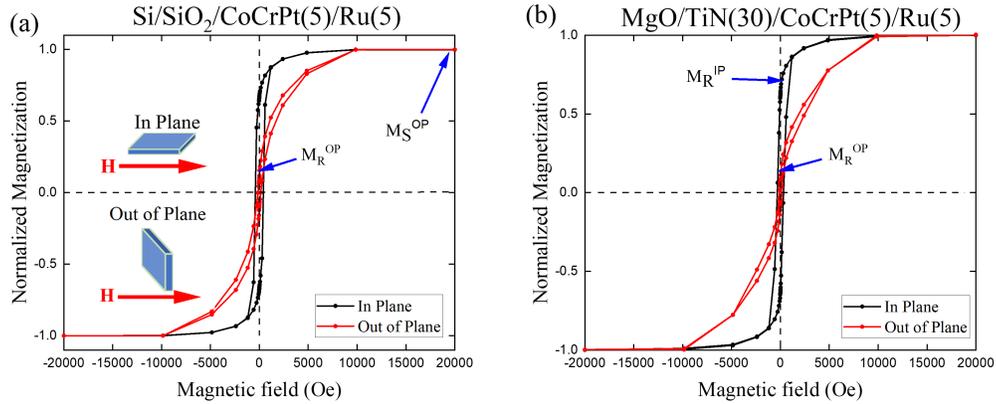

Figure 1. M-H hysteresis loops of $Co_{70}Cr_{18}Pt_{12}$ films measured parallel to and perpendicular to the thin film plane (see insert): (a) growth on $Si/SiO_2$ substrates: the arrows indicate the magnitudes of $M_R^{OP}$ and $M_S^{OP}$; (b) growth on MgO/TiN(30nm) substrates, the arrows indicate the magnitudes of $M_R^{OP}$ and $M_R^{IP}$. Comparison of the hysteresis loops clearly indicates that the magnetization orientation of $Co_{70}Cr_{18}Pt_{12}$ directly grown on these substrates is in the plane of the thin film.

## 3.2 $Co_{70}Cr_{18}Pt_{12}$ thin film growth on $Si/SiO_2$ and MgO/TiN substrates coated with dual Ta/Ru seed interlayers

Deposition of $Co_{70}Cr_{18}Pt_{12}$ on Ta(5nm)/Ru(10nm) dual layers deposited on both $Si/SiO_2$ and MgO/TiN substrates resulted in PMA as evidenced in Figs. 2(a) and 2(b). The S* values and $M_R^{OP}/M_R^{IP}$ ratios are 0.933 & 19.10 and 0.878 and 7.46 for growth on $Si/SiO_2$ and MgO/TiN substrates respectively. This is indicative that the dual layers are effective in inducing growth of the hcp-Co alloy with is basal (0002) plane oriented parallel to the substrate plane, thereby aligning the c-axis preferentially out-of-plane. Comparison of the magnetic figures of merit for the two substrates, shows that the degree of c-axis alignment (crystallographic texture) is higher for growth on $Si/SiO_2$/Ta/Ru substrates. This is likely due to the presence of titanium oxides or titanium oxynitride phases on the TiN surface formed, as previously described, when MgO/TiN samples are exposed to ambient air. These secondary phases can be expected to alter the surface energy and wettability of the TiN interface resulting in degradation of the microstructural properties of the Ta/Ru dual layers. As previously discussed, for strong spin-photon coupling, 15nm thick dual layers are unacceptable. Thus, the focus of this work is identifying nm thick interlayers that promote c-axis orientation out-of-plane in $Co_{70}Cr_{18}Pt_{12}$.

The c-axis in hcp-Co alloys is perpendicular to the (0002) basal plane, therefore XRD can be used to readily confirm the required crystalline growth orientation for PMA development. In the case of thin films grown on MgO substrates, the hcp-Co alloy (0002) reflection in normal incidence x-ray scans, exactly overlaps with that of the (002) planes of MgO. Therefore, in this work we studied the evolution of c-axis out-of-plane orientation in $Co_{70}Cr_{18}Pt_{12}$ vs. the composition and thickness of the investigated interlayers on structures grown on $Si/SiO_2$ substrates. Identical samples were simultaneously grown on MgO/TiN substrates and magnetic properties are measured and compared to determine the efficacy of the seed layers on the development of PMA in $Co_{70}Cr_{18}Pt_{12}$ grown on both MgO/TiN and $Si/SiO_2$ substrates. To correlate structural and magnetic properties, 2θ-ω XRD scans and rocking curve measurements in $Co_{70}Cr_{18}Pt_{12}$, $(Co_{70}Cr_{18}Pt_{12})_xTa_y$ and Ta/Ru dual-layers were conducted. The full width at half maximum (FWHM) of the rocking curve measurements of the CoCrPt (0002) reflection is an indicator of the degree of alignment of the magnetic axis perpendicular to the thin film plane.

For example, Fig. 2(c) provides the rocking curve measurement for a $Co_{70}Cr_{18}Pt_{12}$ sample grown on a $Si/SiO_2/Ta(5nm)/Ru(10nm)$ structure. A FWHM value of 5.92° is measured indicating a high degree of crystalline texture. In comparison, growth of $Co_{70}Cr_{18}Pt_{12}$ on $Si/SiO_2/[(Co_{70}Cr_{18}Pt_{12})_{50}Ta_{50}](10nm)$ yielded a FWHM value of 8.42° (Fig. 2(d)) indicative of a higher degree of dispersion of the c-axis.

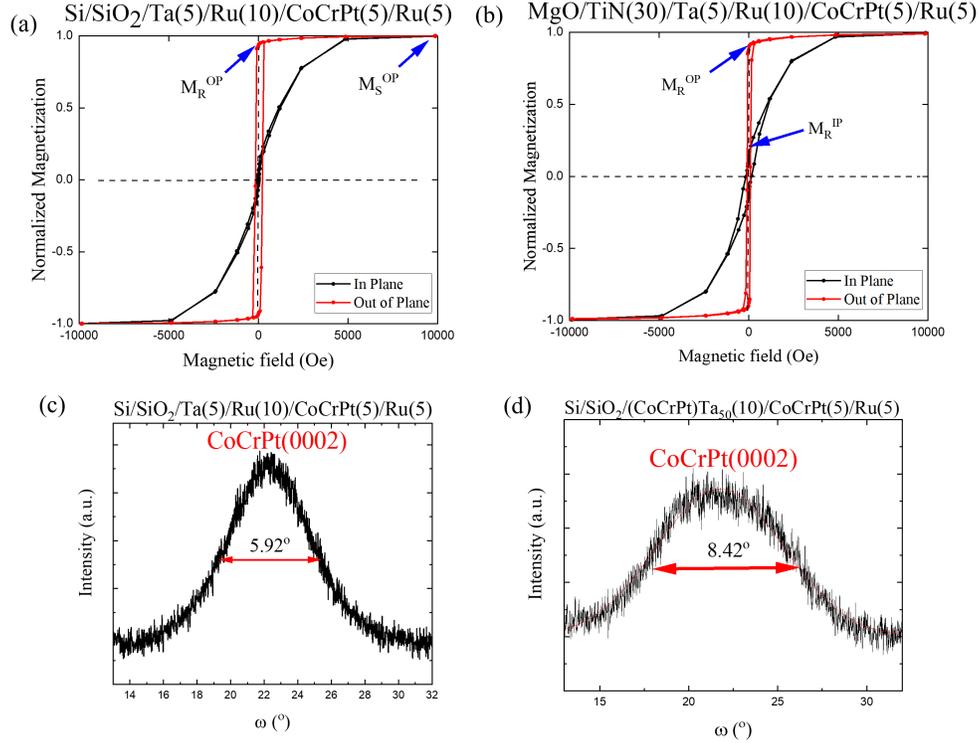

Figure 2. M-H hysteresis loops for $Co_{70}Cr_{18}Pt_{12}$ films deposited on: (a) $Si/SiO_2(300nm)/Ta(5nm)/Ru(10nm)$. The development of strong PMA in this film indicates that the c-axis aligns perpendicular to the thin film; (b) PMA is also observed for growth on $MgO/TiN(30nm/Ta(5nm)/Ru(10nm)$. However, as observed, the larger value of $M_R^{IP}$ in this case indicates significant in-plane orientation of the c-axis (c) Rocking curve measurement of the CoCrPt (0002) reflection for growth on $Ta(5nm)/Ru(5nm)$ dual layers. (d) Rocking curve measurement of the CoCrPt (0002) reflection for growth on $(CoCrPt)Ta_{50}(1nm)$ seed layer.

### 3.3 $Co_{70}Cr_{18}Pt_{12}$ growth on $Si/SiO_2$ (MgO/TiN)/$(Co_{70}Cr_{18}Pt_{12})_xTa_y$ seed layers

The $(Co_{70}Cr_{18}Pt_{12})_xTa_y$ alloy system is a versatile materials platform to control the crystalline growth of thin films subsequently deposited, on account of the multiple structural transformations that are induced by Ta additions. Incorporation of Ta into the hcp unit cell of CoCrPt alloys increments its lattice parameters on account of its larger atomic radius compared to the other atomic constituents: Ta=0.147nm, Pt=0.138nm, Cr=0.128nm, Co=0.125nm. Furthermore, Ta-doping in CoCrPt has been reported to induce a crystalline to amorphous transformation at Ta-doping levels of ~ 10%[28]. An additional consequence of incrementing the atomic concentration of larger atoms into the hcp-Co unit cell is a structural transformation from hcp to fcc as reported by Toney *et. al*[29] for the case of Pt substitutional doping of CoCrPtB alloys. Said structural transformation is driven by strain-relaxation and was observed for a Pt content above 35%. In addition to structural transformations, Ta doping of CoCrPt reduces is saturation magnetization. The following values of the $Co_{70}Cr_{18}Pt_{12}$ saturation moment (emu/cc) were measured as a function of Ta doping: 477, 66, 5 and 1 for Ta (%)= 0, 20, 40 and 50 respectively. To study the efficacy of Ta-doped CoCrPt alloy seed layers as interlayers in magneto-photonic structures to promote $Co_{70}Cr_{18}Pt_{12}$ (0002) fiber texture in the film normal direction, CoCrPt thin films were grown on seed layers of varying composition and thicknesses.

Table 1 summarizes magnetic FOM for $Co_{70}Cr_{18}Pt_{12}$ (5nm) thin films deposited on $Si/SiO_2$(300nm) and MgO/TiN(30nm) substrates overcoated with 1nm thick interlayers of $(Co_{70}Cr_{18}Pt_{12})_xTa_y$ with y=0, 20, 30, 40%. The table includes also for reference, FOM values for $Co_{70}Cr_{18}Pt_{12}$ growth on Ta(5nm)/Ru(10nm) bilayers. $M_R^{OP}/M_R^{IP}$, the ratio of the out-of-plane magnetization over that measured in the in-plane direction, in the absence of an external applied magnetic field, directly measures the degree of PMA in CoCrPt. This FOM can be observed to strongly increment with the Ta content in the seed layer. Comparison of FOM values for growth on $Si/SiO_2$(300nm) and MgO/TiN(30nm) substrates indicates that a higher degree of PMA is attained on $Si/SiO_2$(300nm) substrates, we ascribe this effect to the difference in surface energy between the $SiO_2$ and $TiO_xN_y$ interfaces and its effect on the structural evolution of the seed layers. It is noteworthy that the 1nm thick $(Co_{70}Cr_{18}Pt_{12})_{60}Ta_{40}$ seed layer is more effective than the 15nm Ta/Ru bilayer in promoting PMA in CoCrPt grown on MgO/TiN substrates. $S^*=M_R^{OP}/M_S^{OP}$, the ratio of the remanent out-of-plane moment to that obtained under a strong magnetic field, is an indicator of the magnetic domain structure of the thin film at remanence. Microstructural properties such as grain size, degree of crystalline texture and local compositional variations control this. The results of Table 1 are indicative that growth on $Si/SiO_2$ yields more homogeneous microstructures than in MgO/TiN at lower Ta-content in the seed layer. $S^*$ is highest for the growth on the bilayer structures, consistent with the lower FWHM of the rocking curve of the CoCrPt (0002) reflection measured in these structures. Noteworthy is also the observation that the FOMs for the case of growth on MgO/TiN decrease when the Ta content in the seed layer is 50%. Table 1 also presents the measured saturation magnetization ($M_S$) of each corresponding sample. It is evident that the magnetization of the samples grown on MgO/TiN substrates are consistently higher than when grown on $Si/SiO_2$, which could be attributed to a thinner magnetic dead layer between the TiN/(CoCrPt)Ta, TiN/CoCrPt, or TiN/Ta interfaces compared to $SiO_2$.

**Table 1 Magnetic properties of CoCrPt films grown on ultrathin $(CoCrPt)_xTa_y$ seed layers**

| | $Si/SiO_2$ | | | MgO/TiN | | |
|---|---|---|---|---|---|---|
| Seed layer | $M_R^{OP}/M_R^{IP}$ | $S^* = M_R^{OP}/M_S^{OP}$ | $M_S$ (emu/cc) | $M_R^{OP}/M_R^{IP}$ | $S^* = M_R^{OP}/M_S^{OP}$ | $M_S$ (emu/cc) |
| No seed | 0.16 | 0.13 | 480 | 0.11 | 0.01 | 430 |
| $(Co_{70}Cr_{18}Pt_{12})_{80}Ta_{20}$(1nm) | 4.6 | 0.89 | 530 | 2.8 | 0.73 | 640 |
| $(Co_{70}Cr_{18}Pt_{12})_{70}Ta_{30}$(1nm) | 7.2 | 0.84 | 460 | 3.9 | 0.76 | 590 |
| $(Co_{70}Cr_{18}Pt_{12})_{60}Ta_{40}$(1nm) | 11 | 0.89 | 460 | 9.4 | 0.82 | 570 |
| $(Co_{70}Cr_{18}Pt_{12})_{50}Ta_{50}$(1nm) | 19 | 0.87 | 430 | 7.9 | 0.75 | 510 |
| **Ta(5nm)/Ru(10m)** | **19** | **0.93** | **480** | **7.5** | **0.88** | **550** |

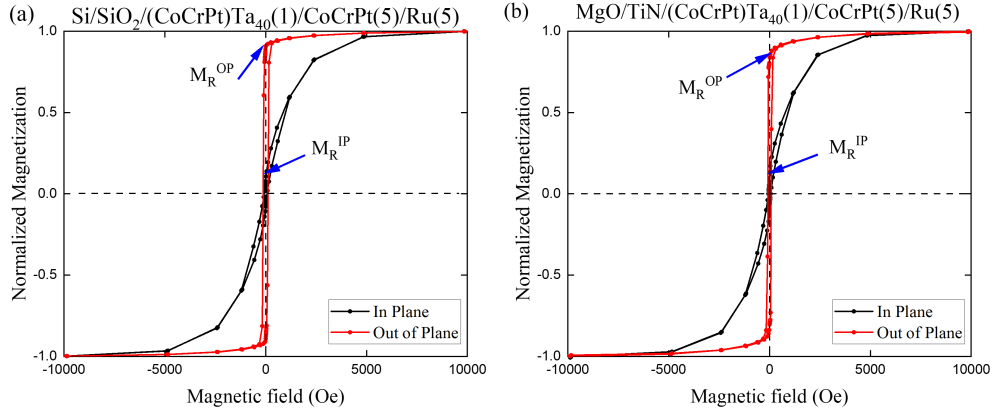

Figure 3. M-H hysteresis loops for $Co_{70}Cr_{18}Pt_{12}$ (5nm) films deposited on $(Co_{70}Cr_{18}Pt_{12})_{60}Ta_{40}$ (1nm) interlayers on: (a) Si/SiO$_2$(300nm) and (b) MgO/TiN(30nm). The development of strong PMA indicates that the c-axis aligns perpendicular to the thin film.

Representative magnetic hysteresis loops for $Co_{70}Cr_{18}Pt_{12}$ (5nm) thin films grown on $(Co_{70}Cr_{18}Pt_{12})_{60}Ta_{40}$ (1nm) interlayers deposited on Si/SiO$_2$ (300nm) and on MgO/TiN (30 nm) substrates are presented in Fig. 3. It is apparent that the ultrathin interlayers seed the required crystallographic growth in $Co_{70}Cr_{18}Pt_{12}$ (5nm) for c-axis alignment perpendicular to the thin film plane.

*3.4 Optical properties of TiN and TiN/CoCrPtTa/CoCrPt/Ru magneto-photonic stacks*

The optical properties of photonic 30nm TiN thin films and of the multilayer magneto-photonic stacks (TiN/(CoCrPt)Ta/CoCrPt/Ru) were measured using variable angle spectroscopic ellipsometry (J. A. Woollam ellipsometer) at angles of 50° and 70° for wavelengths from 500nm to 2000nm. For the TiN sample, a Drude – Lorentz model was used to fit the measurements,

$$\varepsilon(\omega) = \varepsilon_\infty - \frac{\omega_p^2}{\omega^2 + i\Gamma_D \omega} + \sum_L \frac{f_L \omega_L^2}{\omega_L^2 - \omega^2 - i\Gamma_L \omega} \qquad (1)$$

where $\varepsilon_\infty$ is the permittivity at high frequency, $\omega_p$ is the plasma frequency, $f_L$ is the strength of the Lorentz oscillator, $\omega_L$ is the resonant frequency of the Lorentz oscillator, and $\Gamma_L$ and $\Gamma_D$ are the damping of the oscillators. The Drude term captures the optical response of the free carriers, while the Lorentz term accounts for interband transitions. The extracted complex permittivity is shown in Fig. 4a.

In the case for the multilayer stack, the layers $(Co_{70}Cr_{18}Pt_{12})_{60}Ta_{40}$(1nm), $Co_{70}Cr_{18}Pt_{12}$(5nm), and Ru(5nm) were treated as a single effective layer on top of the measured TiN film and fitted using a Drude – Lorentz model, as well. Since each layer on top of the TiN is a lossy material, we assume that this model provides a valid estimate of the effective complex permittivity of the multilayer stack. The complex permittivity of the single effective layer is plotted in Fig. 4b.

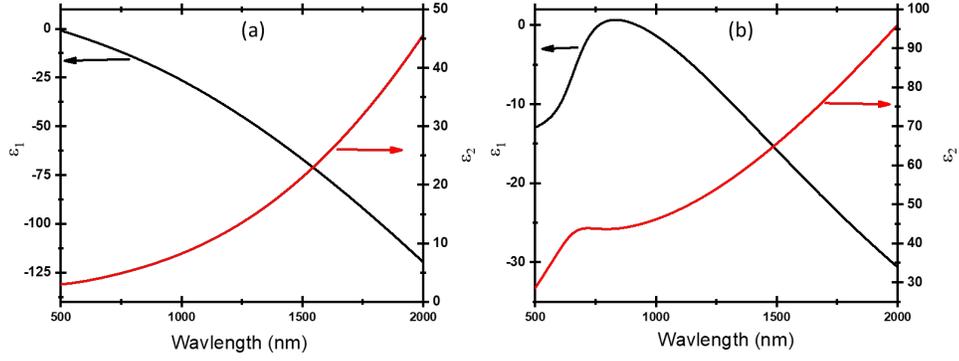

Figure 4. Complex permittivity of (a) 30 nm TiN on MgO and (b) (CoCrPt)Ta/CoCrPt/Ru on TiN

To assess the plasmonic performance of our materials, we have calculated the surface plasmon polariton (SPP) dispersion curves for the TiN film and the multilayer system. The SPP dispersion relation at the air/TiN interface is given by,

$$\beta = k_0 \sqrt{\frac{\varepsilon_{TiN}}{\varepsilon_{TiN} + 1}} \tag{2}$$

where $\beta$ is the complex propagation constant, $k_0$ is the wave vector of the propagating wave in vacuum, $\varepsilon_{TiN}$ is the complex permittivity of TiN.

The SPP dispersion relation for the multilayer system can be determined by solving the following expression [30],

$$e^{-4k_1 a} = \frac{\left(\frac{k_1}{\varepsilon_1} + \frac{k_2}{\varepsilon_2}\right)\left(\frac{k_1}{\varepsilon_1} + \frac{k_3}{\varepsilon_3}\right)}{\left(\frac{k_1}{\varepsilon_1} - \frac{k_2}{\varepsilon_2}\right)\left(\frac{k_1}{\varepsilon_1} - \frac{k_3}{\varepsilon_3}\right)} \tag{3}$$

where $a$ is half the thickness of the (CoCrPt)Ta/CoCrPt/Ru stack and $\varepsilon_1, \varepsilon_2$ and $\varepsilon_3$ are the permittivities of the (CoCrPt)Ta/CoCrPt/Ru stack, TiN and air, respectively (Fig. 5a) . $k_i$ ($i$ = 1, 2, 3) is given by applying continuous solutions that fulfill the wave equation,

$$k_i^2 = \beta^2 - k_0^2 \varepsilon_i \tag{4}$$

The dispersion curves for the bare TiN film and the TiN film with the (CoCrPt)Ta/CoCrPt/Ru stack are shown in Fig. 5b. We note that the dispersion curve for the multilayer magnetic stack is an approximation and not the exact solution since the optical properties of each individual layer are not extracted. Since each of the layers on top of the TiN have high losses and small thicknesses, we assume that the SPP decay in each of the layers may not widely vary and the effective permittivity of the multilayer stack provides a reasonable approximation of the overall SPP dispersion.

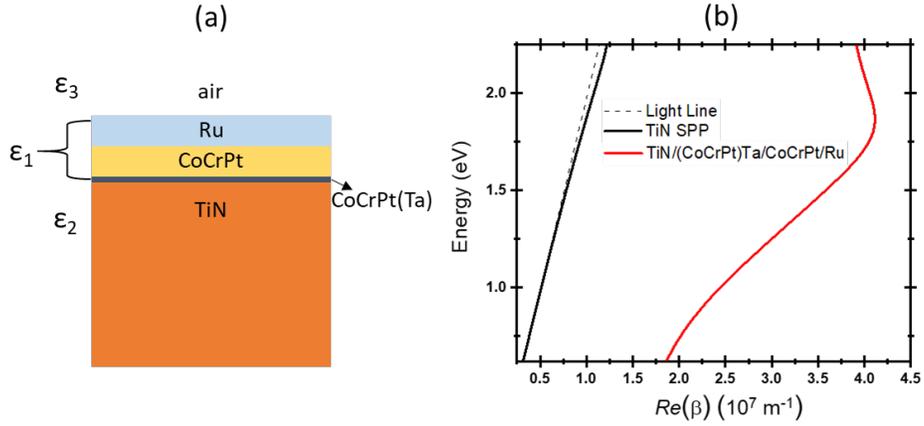

Figure 5. (a) Schematic of sample structure and corresponding layer permittivities per Eq. 3. (b) SPP dispersion curves for TiN/air and TiN/[(CoCrPt)Ta/CoCrPt/Ru]/air. The light line is plotted as the dashed line.

We calculate the figure of merit (FOM) for both the bare TiN film and the multilayer system as $FOM = \frac{Re(\beta)}{Im(\beta)}$ (Fig. 6). Because of the addition of lossy materials on top of the low-loss TiN, the figure of merit is decreased in the multilayer system compared to the bare TiN film.

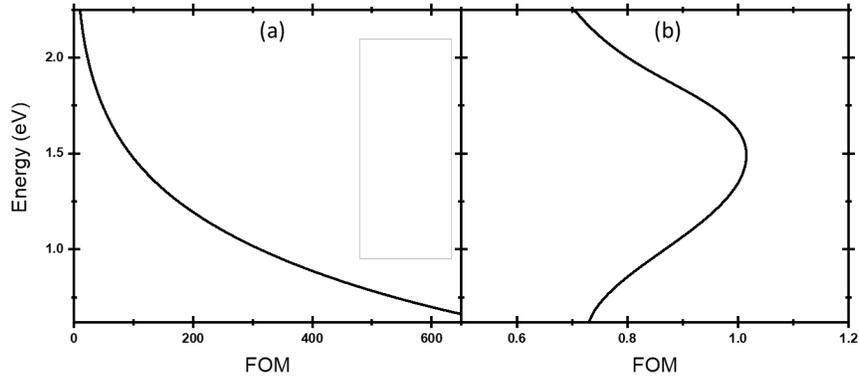

Figure 6. Figure of merit for (a) TiN/air and (b) [(CoCrPt)Ta/CoCrPt/Ru]/air

## *4 Discussion*

To understand the evolution of the perpendicular magnetic anisotropy in $Co_{70}Cr_{18}Pt_{12}$ on the Ta content of the $(Co_{70}Cr_{18}Pt_{12})_xTa_y$ interlayers, the microstructural properties of the interlayers as a function of Ta content were determined by XRD. To this effect, we grew a series of 10 nm thick $(Co_{70}Cr_{18}Pt_{12})_xTa_y$ seed layers with y=0, 20, 30, 40 and 50% on Si/SiO$_2$ (300nm) substrates. The choice of the substrate was driven, as discussed previously, by our instrumental inability to resolve the full overlap between the Co (0002) and the MgO (002) XRD diffraction peaks. Furthermore, 10nm thick seed layers were chosen as our XRD system does not have the sensitivity to enable us measurements on 1nm seed layers.

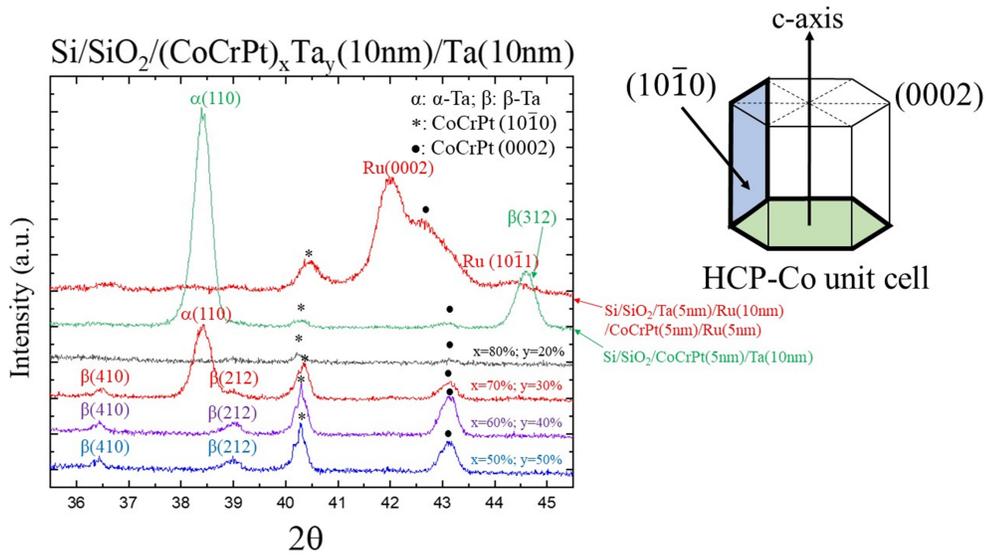

Figure 7. 2Θ-ω XRD scans of $(CoCrPt)_xTa_y$ seed layers (lower 4 spectra) deposited on $Si/SiO_2$ and capped with Ta(10nm). Top spectra: CoCrPt(5nm) grown on $Si/SiO_2$ /Ta(5nm)/Ru(10nm) bilayers and capped with Ru(5nm) and growth of CoCrPt (5nm) directly on $Si/SiO_2$ and capped with Ta(10nm). Legends: $\alpha = \alpha$-Ta, $\beta = \beta$-Ta, ∗= hcp ($10\bar{1}0$) diffraction; •: hcp (0002) diffraction. The insert shows the hcp unit cell with the crystallographic planes of interest identified. The magnetic axis in hcp Co-alloys aligns parallel to the ($10\bar{1}0$) plane.

Fig. 7 presents XRD spectra for the 10nm $(Co_{70}Cr_{18}Pt_{12})_xTa_y$ seed layer series grown on $Si/SiO_2$ substrates. For comparison, those corresponding to a 5nm $Co_{70}Cr_{18}Pt_{12}$ thin film capped with Ta(10nm) is provided, as well as for the 5nm $Co_{70}Cr_{18}Pt_{12}$ grown on Ta(5nm)/Ru(10nm) seed layers and capped with Ru(5nm). The spectra have shifted for clarity. The insert depicting the hcp-Co unit cell and the crystallographic planes relevant to the results discussion are indicated. The magnetic moment in hcp-Co alloys aligns parallel to the c-axis and the ($10\bar{1}0$) plane. The spectrum for the CoCrPt grown on dual Ta/Ru bilayers exhibits hcp-CoCrPt diffractions corresponding to the ($10\bar{1}0$) and (0002) planes. The intensity of the (0002) peak is significantly higher than that of the ($10\bar{1}0$), validating the strong PMA observed in this film (refer to Fig. 1 and Table 1). Diffraction peaks corresponding to the Ru (0002) underlayer and capping layer are also evident. The Ru capping layer exhibits ($10\bar{1}1$) orientation. As can be seen in this spectrum, there is significant overlap between the Ru and CoCrPt (0002) diffraction peaks, therefore, to facilitate the structural analysis, we replaced the Ru protective layer with Ta. The XRD spectrum for direct growth of CoCrPt on $Si/SiO_2$ exhibits diffractions corresponding to the ($10\bar{1}0$) and (0002) planes (second spectrum from the top). The intensity of the ($10\bar{1}0$) reflection being the stronger one and therefore the c-axis for this sample is mostly in-plane. The strongest peaks observed in this spectrum correspond to the α-Ta(110) and β-Ta(312) phases of the Ta cap. The spectrum for the $(Co_{70}Cr_{18}Pt_{12})_{80}Ta_{20}$ alloy is essentially that of an amorphous material. Note that the 10nm Ta overlayer exhibits also no crystalline peaks, indicative that Ta grows amorphous on the amorphous CoCrPtTa seed layer. This is consistent with the phase transformation in CoCrPt reported by Lee and Kim [23] when the alloy is doped with >6% Ta. Beginning with a Ta content of 30% in the seed layer, diffraction peaks corresponding to ($10\bar{1}0$) and (0002) planes of an hcp unit cell can be observed. Additional peaks in these spectra originate from crystalline planes of the Ta capping layer. Note the structural changes of the Ta capping layer when grown on different Ta-containing seed layers: the α-Ta and as β-Ta phases are present when the Ta content is 30%, but for higher values, Ta grows only as β-Ta. We note also the significant shift of the (0002) diffraction peak positions for the $Co_{70}Cr_{18}Pt_{12}$ alloy grown on Ta(5nm)/Ru(10nm) dual layers (42.42°) and that of the crystalline $(Co_{70}Cr_{18}Pt_{12})_{x>30}Ta_{y<70}$ (~43.1°) seed layers grown directly on Si/SiO$_2$. This is explained by the fact that the (0002) plane of CoCrPt grown on Ru underlayers forms, through pseudo-epitaxy, with comparable lattice spacing as the Ru (0002) plane (42°). Whereas for the case of the crystalline $(Co_{70}Cr_{18}Pt_{12})_{x>30}Ta_{y<70}$ seed layers, the (0002) plane adopts a spacing that minimizes the unit-cell lattice strain energy without epitaxial constraints imposed by amorphous SiO$_2$.

To correlate the structural changes observed in the Ta-doped CoCrPt seed layers as a function of Ta-content with the measured magnetic FOM of the CoCrPt 5nm thin films grown on them, we use the results presented in Figs. 1-3, 7 and in Table 1. In our analysis we make the assumption that the structural changes observed in the 10nm thick CoCrPtTa seed layers are also present in the case of the ultrathin 1nm seed layers. Whereas we recognize that in the ultrathin film regime additional factors such as enhanced lattice strain, structural defects and interface chemistry may be present, the magnetic changes we observe in both set of seed layers are similar. Fig. 1 and Table 1 indicate that CoCrPt grown directly on Si/SiO$_2$ and MgO/TiN substrates have their magnetization orientation in the film plane, consistent with the XRD results of Fig. 7. The increase in PMA in CoCrPt when grown on CoCrPtTa alloys with incrementing amounts of Ta as evidenced in Fig. 3 and Table 1, is correlated to the progressive increment of the intensity of the hcp-CoCrPtTa (0002) diffraction peaks (Fig. 7). As in the case of growth on Ta/Ru dual layers, the CoCrPt grows on seed layers with close (0002) lattice match, thereby enabling c-axis fiber texture resulting in the perpendicular orientation of the magnetic axis. The XRD results can explain also subtle effects observed in Table 1. We note that $S^* = M_R^{OP}/M_S^{OP}$ that is indicative of the degree of texture, maximizes for Ta=40% and slightly decreases when the Ta-content increments to 50%. This is consistent with the analysis of the integrated peak intensity ratio $[(0002)/(10\bar{1}0)]$ dependence on the Ta-content of the seed layers: the ratio is 0.62, 1.26 and 1.05 for 30, 40 and 50% Ta content respectively.

We note that for the case of the hybrid magneto-photonic structures proposed by Dutta *et al.*[11] for ultrafast magnetic switching, said metallic interlayer thicknesses between the plasmonic and magnetic material would introduce significant optical losses, precluding the strong photon-spin coupling required for generating enhanced OMF. To circumvent this problem, in this work we report on the synthesis of few atomic-thick interlayers capable of seeding PMA in CoCrPt thin films grown on plasmonic TiN. Structural and magnetic property measurements are employed to characterize the hybrid structures.

As mentioned in the introduction, reduction of the metallic interlayer thickness between the plasmonic and magnetic material to minimize optical losses is critical. This is obviated by the ellipsometry results provided in the supplementary information. Effective medium modeling was employed to quantify the lossy behavior of thick interlayers. The ellipsometry measurements of MgO/TiN(30)/CoCrPt$_{60}$Ta$_{40}$/CoCrPt(5)/Ru(5) multilayer film stacks were analyzed treating the top CoCrPt$_{60}$Ta$_{40}$/CoCrPt(5)/Ru(5) films as an effective layer which was fitted using a Drude-Lorentz model. The results show a highly lossy behavior for this thick multilayer structure. As the constituent layers of the stack are all metallic and highly lossy, we can conclude that we benefit from thinner seed layer thicknesses when considering E-field amplitudes and consequently the potential OMF enhancement.

## 4. Conclusions

A magneto-plasmonic hybrid material structure has been demonstrated that can potentially be used for AOS of the magnetization orientation in nanoscale structures in fs time scales. Utilizing judiciously chosen quaternary CoCrPtTa alloy compositions, we are able to control the growth orientation of the magnetic layer to develop strong magnetic perpendicular anisotropy in magnetic alloys grown on TiN refractory plasmonic materials. This has been achieved with the aid of 1nm thick interlayers between the plasmonic and the magnetic material, thereby enabling us to achieve two key requirements for magneto-photonic AOS: strong photon-spin coupling and PMA.

The crystallographic properties of the $(CoCrPt)_x Ta_y$ interlayers were controlled by changing the Ta content. Increasing Ta-doping of CoCrPt resulted initially in a crystalline to amorphous phase transformations, and the subsequent development of an hcp-crystalline structure characterized by the coexistence of $(10\bar{1}0)$ and $(0002)$ crystalline orientations. The desired $(0002)$ orientation for PMA development was found to depend on the Ta-content.

Whereas our structural analysis was limited to normal incidence diffraction and thick seed layers, a wealth of new information can be obtained by utilizing synchrotron sources for higher resolution, ultra-sensitive measurements in normal and grazing incidence as well as EXAFS and XAS techniques to obtain an atomic description of the local environment. This information will be invaluable to suppress the formation of the $(10\bar{1}0)$ hcp orientation, thereby significantly incrementing even further PMA with potentially atomic-thick seed layers.

## 5. Funding and acknowledgements

The authors would like to acknowledge support from the Office of Naval Research Grant No. N00014-16-1-3003 and from Purdue University School of Materials Engineering for support for Alan Chu.

## 6. Disclosures

The authors declare that there are no conflicts of interest related to this article.